\pdfoutput=1
\documentclass[conference,a4paper,11pt]{IEEEtran}
\usepackage[keeplastbox]{flushend}
\usepackage[latin1]{inputenc}
\usepackage{times,amsmath}
\usepackage{amsfonts}
\usepackage{pstool}
\usepackage{subfigure}
\usepackage{multirow}
\usepackage{multicol}
\usepackage{enumerate}
\usepackage{graphicx}
\usepackage{mathtools, cuted}
\usepackage{MnSymbol}
\usepackage{stfloats}
\usepackage[table]{xcolor}
\usepackage{nohyperref}
\usepackage{algorithm,algorithmic}
\usepackage{bigints}
\usepackage{amsmath}
\begin{document}
\title{Preventing Identity Attacks in RFID Backscatter Communication Systems: A Physical-Layer Approach}
\author{
\IEEEauthorblockN{Ahsan Mehmood$^\ast$, Waqas\ Aman$^{\ast \dagger} $, M.\ Mahboob\ Ur\ Rahman$^\ast$, M.\ A.\ Imran$^\dagger$ and Qammer\ H.\ Abbasi$^\dagger$\\
$^\ast$Electrical engineering department, Information Technology University, Lahore 54000, Pakistan \\
$^ { \dagger}$Department of Electronics and Nano engineering, University of Glasgow, Glasgow, G12 8QQ, UK \\
$^\ast$\{msee19009,waqas.aman,mahboob.rahman\}@itu.edu.pk,\\ $^\dagger$\{muhammad.imran,qammer.abbasi\}@glasgow.ac.uk
}
}
\maketitle

%\author{\IEEEauthorblockN{Ahsan\ Mahmood$^\ast$, Waqas Aman$^\ast$ $^\dager $, Muhammad\ Mahboob\ Ur\ Rahman$^\ast$, and Octavia\ A.\ Dobre$^\ddager$} \\
%Electrical engineering Department, Information Technology University, Lahore 54000, Pakistan \\
%Department of Electronics and Nano Engineering, University of Glasgow, Glasgow, G12 8QQ, UK \\
%Department of Electrical and Computer Engineering, Memorial University, St.  John's, NL A1B 3X5, Canada}

%\long\def\symbolfootnote[#1]#2{\begi%ngroup%
%\def\thefootnote{\fnsymbol{footnote}%}\footnote[#1]{#2}\endgroup}
%\symbolfootnote[0]
%    {\hrulefill \\
 %Ahsan Mahmood and Muhammad Mahboob Ur Rahman are with the Electrical engineering Department, Information Technology University, Lahore 54000, Pakistan (\{ahsan.mahmood, mahboob.rahman\}@itu.edu.pk). \\
%Octavia A. Dobre is with the Department of Electrical and Computer Engineering, Memorial University, St.  John's, NL A1B 3X5, Canada (odobre@mun.ca).}
\begin{abstract}

This work considers identity attack on a radio-frequency identification (RFID)-based backscatter communication system. Specifically, we consider a single-reader, single-tag RFID system whereby the reader and the tag undergo two-way signaling which enables the reader to extract the tag ID in order to authenticate the legitimate tag (L-tag). We then consider a scenario whereby a malicious tag (M-tag)---having the same ID as the L-tag programmed in its memory by a wizard---attempts to deceive the reader by pretending to be the L-tag. To this end, we counter the identity attack by exploiting the non-reciprocity of the end-to-end channel (i.e., the residual channel) between the reader and the tag as the fingerprint of the tag. The passive nature of the tag(s) (and thus, lack of any computational platform at the tag) implies that the proposed light-weight physical-layer authentication method is implemented at the reader. To be concrete, in our proposed scheme, the reader acquires the raw data via two-way (challenge-response) message exchange mechanism, does least-squares estimation to extract the fingerprint, and does binary hypothesis testing to do authentication. We also provide closed-form expressions for the two error probabilities  of interest (i.e., false alarm and missed detection). Simulation results attest to the efficacy of the proposed method. 

\end{abstract}

\begin{IEEEkeywords}
backscatter communication, authentication, identity attacks, non-reciprocal hardware, transmitter identification, intrusion detection 
\end{IEEEkeywords}
\section{Introduction}
In backscatter communication systems, the reader/interrogator sends out a continuous-wave signal, while the tag utilizes passive reflection or load modulation of incident radio frequency wave to respond back to the reader. This two-way (challenge-response) message exchange helps the reader extract useful information  from the intended tag\cite{Niu:JCIN:2019}. Backscatter communication systems are widely utilized in a number of application scenarios, e.g., transport and  medical industries, access control, smart parking and smart grids, to name a few \cite{Van:CST:2018}. The increasing demand of high data rates in backscatter communication systems has prompted the researchers to build full-duplex backscatter communication systems \cite{6068681}. In a typical single-reader, single-tag, full-duplex backscatter communication system, the reader transmits and receives the radio signal simultaneously whereas  the tag load modulates and backscatters the radio signal transmitted by the reader. \\
The broadcast nature of backscatter communication makes it vulnerable to many attacks by adversaries. To counter such attacks in many wireless communication systems, cryptographic measures have been widely used. But, there are limitations of these crypto-based measures. One main limitation is the dependency on the shared secret or shared secret key among legal nodes. Recently, authors in \cite{gidney:arXiv:2019} reported that the integrity of crypto-based measures is at high risk due to advances in quantum computing, and thus, crypto-based measures are quantum insecure. In this regard, physical layer security (PLS) which exploits the unique characteristics of the propagation medium \cite{Aman:WCNC:2016}, \cite{Aman:ETT:2018} is a promising approach to complement the crypto-based schemes at the higher layers of the protocol stack.

One essential ingredient of the PLS is Physical layer authentication (PLA)---basically a tool which a receiver could utilize to verify the identities of the transmit nodes. PLA has received considerable attention of the researchers due to its light-weight implementation and robust nature. PLA exploits the unique features of the propagation medium which are nearly impossible to clone unless and until the malicious node is co-located with the legitimate node whose probability is almost zero in practice. To date, there are various features reported for physical layer authentication, e.g., received signal strength \cite{Yang:TPDS:2013}, channel impulse response \cite{Aman:VTC:2017, Mahboob:PTL:2020, Aman:UCET:2019} channel frequency response \cite{Xiao:TWC:2008}, carrier frequency offset \cite{Mahboob:Globecome:2014}, pathloss \cite{Mahboob:Access:2017}, \cite{Aman:arXiv:securing}, lack of hardware reciprocity \cite{Mahboob:vtc:2017}, I/Q imbalance \cite{Hao:ICC:2014}  to name a few. More recently, distance, angle-of-arrival and position of the transmit node are reported in \cite{Aman:Access:2018} to thwart the impersonation attack in an underwater acoustic sensor network (UWASN). Last but not the least, \cite{aman:ICC:2020} studies the impact of authentication on effective capacity of an UWASN.\\

Though there are many works reported on the physical layer security in backscatter communication systems, most of them counter the eavesdropping attacks through resource allocation and artificial noise generation. Thus, there are only a handful of works which study physical layer authentication in half-duplex backscatter communication. Specifically, \cite{Luo:ACM:2018} exploited the propagation signatures of the tag, while  \cite{7335078} used analogue fingerprints of the tag. Finally, \cite{12672} reported difference of the radio signal as a fingerprint to authenticate the tag. 
\textit{ In contrast to previous works, this work thwarts identity attacks on RFID-based backscatter communication systems by exploiting residual channel as tag fingerprint in order to carry out physical layer authentication at the reader.}

\textbf{Organization:} 
The rest of the paper is organized as follows.
Section II describes the system model of an RFID-based backscatter communication system that is under identity attack by a malicious tag. Section III presents the proposed physical layer authentication method. Section IV discusses simulation results. Finally, Section V concludes the work. 

\textbf{Notations:} Unless otherwise specified, we use
$(.)^H$ for hermitian, $(.)^T$ for transpose, $(.)^{-1}$ for inverse, uppercase bold face letters for matrix and lowercase bold face letters for vectors (e.g. $\mathbf{X}$ denotes a matrix, $\mathbf{x}$ denotes a vector). Finally, $||\textbf{x}||$ is $l_2$ norm of vector $\textbf{x}$ and  $ \mathcal{CN}$ means complex normal.
\section{System Model \& Background}
\subsection{System Model}
\begin{figure}[h!]
\centering
\includegraphics[scale=0.4]{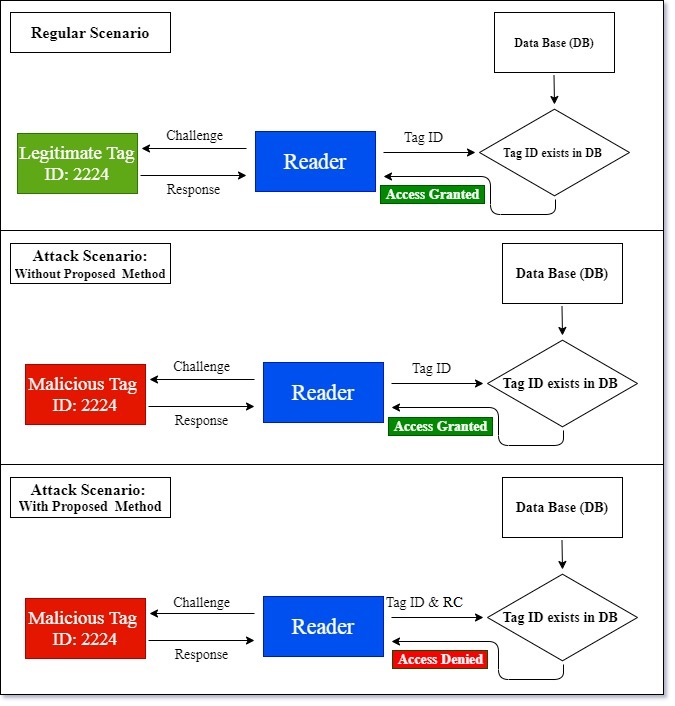}
\caption{System model: consider an RFID-based access control system which fails (i.e., it grants access to the M-tag) when the tag ID of the M-tag is the same as the tag ID of the L-tag. The proposed method counters such identity attacks by doing physical layer authentication using residual channel (RC) as device fingerprint at the reader.} 
\centering
\label{fig:System Model}

\end{figure}

We consider an RFID-based backscatter communication system that comprises a single reader and a single tag. We then consider the situation whereby a malicious tag (M-tag) launches identity attack on the reader by pretending to be the legitimate tag (L-tag), (see Fig. \ref{fig:System Model}). We also learn from Fig. \ref{fig:System Model} that the traditional crypto-based authentication mechanism at the reader fails as long as the M-tag has the same tag ID as that of the L-tag. Therefore, this work proposes to carry out physical layer authentication at the reader whereby the reader measures the fingerprint (residual channel) and compares it against the pre-stored ground truth\footnote{Note that the reader acquires the ground truth/fingerprint of the L-tag on a secure channel offline, and later, estimates the residual channel on the insecure/open channel online.}. Here, it is worth mentioning that this work considers such RFID backscatter communication systems where the tag lies in the far field of the reader.

\subsection{Background: Reciprocity}
The non-reciprocal nature of factory manufactured wireless transceivers can be modeled by their reciprocity parameters (RPs). For a given transceiver, there are two RPs: one for Tx chain and one for Rx chain. Fig. 2 shows the two RPs of the reader ($h^{TX}_R$ and $h^{RX}_R$) and the two RPs of the tag ($h^{TX}_T$ and $h^{RX}_T$). The RPs represent the magnitude and phase distortion caused by the RF chains, and thus, are device dependent and unique. 
%Due to low complexity of RFID tag its receiver and transmitter chain is simple as compared to  reader. 
%The the forward end-to-end channel $h_{TR} = h^{TX}_R h^{RF}_{TR} h_T^{RX}$ and backward end-to-end channel $h_{RT} = h^{TX}_T h^{RF}_{RT} h_T^{TX}$.Due to non-reciprocal nature of RFID reader and tag hardware, the amount of phase and magnitude distortion in communication signals are different during forward propagation and backward propagation.

\begin{figure}[h!]

\centering
\includegraphics[scale=0.8]{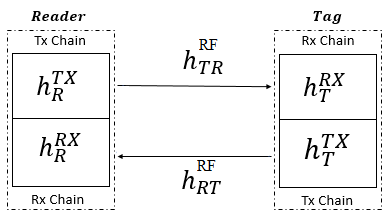}
\caption{Two-way challenge-response signaling}
\centering
\label{fig:Ping pong message exchange}

\end{figure}

\section{Physical Layer Authentication}
The proposed PLA method consists of two steps: (i) feature/fingerprint acquisition using two-way message exchange followed by least-squares estimation of the fingerprint, (ii) binary hypothesis testing for tag identification.

\subsection{Fingerprint Acquisition}
\subsubsection{Two-Way Challenge-Response Signaling}
 We consider "reader-talk-first" communication protocol \cite{6068681}. For the two-way (challenge-response) message exchange, tag follows amplify and forward (AF) relaying mechanism. 
 The reader transmits the challenge message $x_R$ with power $P_R$, and a while later, receives the backscattered response signal from the tag. The challenge message received at the tag at time $n$ is: 
\begin{equation}
     y_{T} [n] = \sqrt{P_R} .x_R [n] .h_{TR}+ \sqrt{P_{SI,T}} z_T [n]
\end{equation}
where $h_{TR} = h_R^{TX} h_{TR}^{RF} h_T^{RX}$ is the end-to-end directional channel from the reader to the tag; $z_T[n]$ is the self-interference signal seen by the tag, and ${P_{SI,T}}$  is the power of self-interference signal at the tag. Inline with previous literature \cite{8302460}, \cite{DBLP:journals/corr/abs-1904-01323}, we do not consider noise in Eq. (1). The response message received at the reader is:
 \begin{equation}
     y_R [n] = \eta h_{RT} y_T [n] + \sqrt{P_{SI,R}} z_R [n]+n_R,
\end{equation}
Where $h_{RT}= h_T^{TX} h_{RT}^{RF} h_R^{RX}$ is the end-to-end directional channel from the tag to the reader and $n_R\sim \mathcal{CN}(0,\sigma_R^2)$ is the noise at the reader. Additionally $\eta$, $z_R [n]$ and $\sqrt{P_{SI,R}}$ represent amplification factor, self-interference signal and power of interference signal at the reader, respectively. 
Equivalently:
\begin{equation}
    \begin{aligned}
        y_R [n] = \sqrt{P_{R}} \eta h_{RT} h_{TR} x_R[n]  +  \sqrt{P_{SI,T}} \eta h_{RT} z_T [n] \\ 
        +  \sqrt{P_{SI,R}} z_R [n] +n_R.
    \end{aligned}
\end{equation}
 Inline with the previous literature \cite{7769255}, we assume that the self-interference signals $\sqrt{P_{SI,T}} \eta h_{RT} z_T\sim \mathcal{CN}(0,\sigma_{R,SI}^2)$ and $\sqrt{P_{SI,R}} z_R\sim \mathcal{CN}(0,\sigma_{T,SI}^2) $. Where $\sigma_{R,SI}^2$ and $\sigma_{T,SI}^2$ is the variance of self interference signal at reader and tag respectively.

Let $n_{RT} = \sqrt{P_{SI,T}} \eta h_{RT} z_T [n] + \sqrt{P_{SI,R}} z_R [n] +n_R$. Then, Eq. (3) can be written as:
\begin{equation}
     y_R [n] = \sqrt{P_{R}} \eta \hat{h}_{RT}  x[n] +n_{RT},
\end{equation} 

Where, $n_{RT} \sim \mathcal{CN}(0,\sigma^2_R + \sigma_{R,SI}^2+\sigma_{T,SI}^2)$, while $\hat{h}_{RT} = h_{TR} h_{RT}$ is the end-to-end residual channel between the reader and the tag.
 \\
%We are using $\hat{h}_{RT}$ as device fingerprint. So, we need to estimate it.
\subsubsection{Least-Square Estimation of Residual Channel}

To estimate the fingerprint $\hat{h}_{RT}$ of the tag, the reader sends $N$ training symbols $\textbf{x}_R = [x_R[1] \ x_R[2] \ ... \ x_R[N]]^T$ during the challenge phase. The tag acts as AF relay and backscatters the amplified signal. Thus, the reader receives $\textbf{y}_R = [y_R[1] \ y_R[2] \  ...\ y_R[N]]^T$ during response phase:
\begin{equation}
    \textbf{y}_R =\hat{h}_{RT}. \hat{\textbf{x}}_R     + \textbf{n}_{RT},
\end{equation}
where $\hat{\textbf{x}}_R = \eta \sqrt{P_R}.\textbf{x}_R $. Then, the least-squares estimate is given as: 
\begin{equation}
   \hat{{{h}}} = (\hat{\textbf{x}}_R^H.\hat{\textbf{x}}_R)^{-1}.\hat{\textbf{x}}_R^H.\textbf{y}_R 
\end{equation}
which implies that $\hat{{{h}}} \sim \mathcal{CN}(\hat{h}_{RT},\sigma^2_{\hat{h}})$.
Here, $\sigma^2 _{\hat{h}} =  \frac{\sigma^2_{RT}}{\eta^2 P_R ||\textbf{x}_R||^2}$ and $\sigma^2_{RT} = \sigma^2_R +  \sigma_{R,SI}^2+\sigma_{T,SI}^2$.

\subsection{Binary Hypothesis Testing}
With the estimate of the tag's fingerprint and perfect ground truth in hand, the reader performs binary hypothesis testing for tag identification. The binary hypothesis test is defined as:
\begin{equation}
	\label{eq:H0H1}
	 \begin{cases} H_0 (\text{L-tag is present}): & \upsilon = \hat{h}_{RT}^{(L)}+\epsilon  \\ 
                   H_1 (\text{M-tag is present}): & \upsilon = \hat{h}_{RT}^{(M)}+\epsilon  \end{cases}
\end{equation}

Here $\epsilon \sim \mathcal{CN}(0,\sigma^2 _{\hat{h}})$ is the estimation error. Then, $\upsilon | H_0 \sim \mathcal{CN}(\hat{h}_{RT}^{(L)},\sigma^2 _{\hat{h}}) $ and $\upsilon | H_1 \sim \mathcal{CN}(\hat{h}_{RT}^{(M)},\sigma^2 _{\hat{h}})$. The test statistic $T$ is given as:
\begin{equation}
        T= {|\upsilon-{\hat{h}_{RT}^{(L)}|}}\underset{H_0}{\overset{H_1}{\gtrless}}{\delta }
\end{equation}
 where $\delta$ is threshold, a design parameter.
     Let $ t = \upsilon-\hat{h}_{RT}^{(L)}$, then $t|H_0 \sim \mathcal{CN}(0,\sigma^2 _{\hat{h}}) $
     and $t|H_1 \sim \mathcal{CN}(\hat{h}_{RT}^{(M)}-\hat{h}_{RT}^{(L)},\sigma^2 _{\hat{h}}) $. Further, $T|H_0 \sim Rayleigh(\sqrt{(\sigma^2 _{\hat{h}}/2})$. Now, the probability of false alarm is:
\begin{equation}
        P_{fa} = Pr(T >\delta|H_0) = exp(\frac{-\delta^2}{\sigma^2 _{\hat{h}}})
\end{equation}
By setting $P_{fa}$ to desired tolerance level, $\delta$ is computed as:
\begin{equation}
        \delta = \sqrt{-ln(P_{fa})\sigma^2 _{\hat{h}}}
\end{equation}
The performance of proposed method can be completely characterized by probability of false alarm and probability of missed detection. Since we are following Neyman-Pearson criterion, the performance of proposed method is solely dependent on success probability of M-tag or probability of missed detection. Since, $T|H_1 \sim Rice(\hat{h}_{RT}^{(M)}-\hat{h}_{RT}^{(L)},\sqrt{\sigma^2 _{\hat{h}}/2})$,
the success probability of M-tag can be expressed as:
\begin{equation}
    P_{md} = Pr(T<\delta|H_1)
\end{equation}
Let $\mu = \hat{h}_{RT}^{(M)}-\hat{h}_{RT}^{(L)} $ and 
$\hat{\sigma} = \sigma^2 _{\hat{h}}/2 $. Then,
\begin{equation}
    P_{md} = 1 - \mathcal{Q}_1(\frac{\mu}{\hat{\sigma}},\frac{\delta}{\hat{\sigma}})
\end{equation}
where $Q_1(.,.)$ is the Marcum Q-function of order $1$.

\section{Simulation Results}
\begin{figure}[h!]
\centering
\includegraphics[width=3.7in]{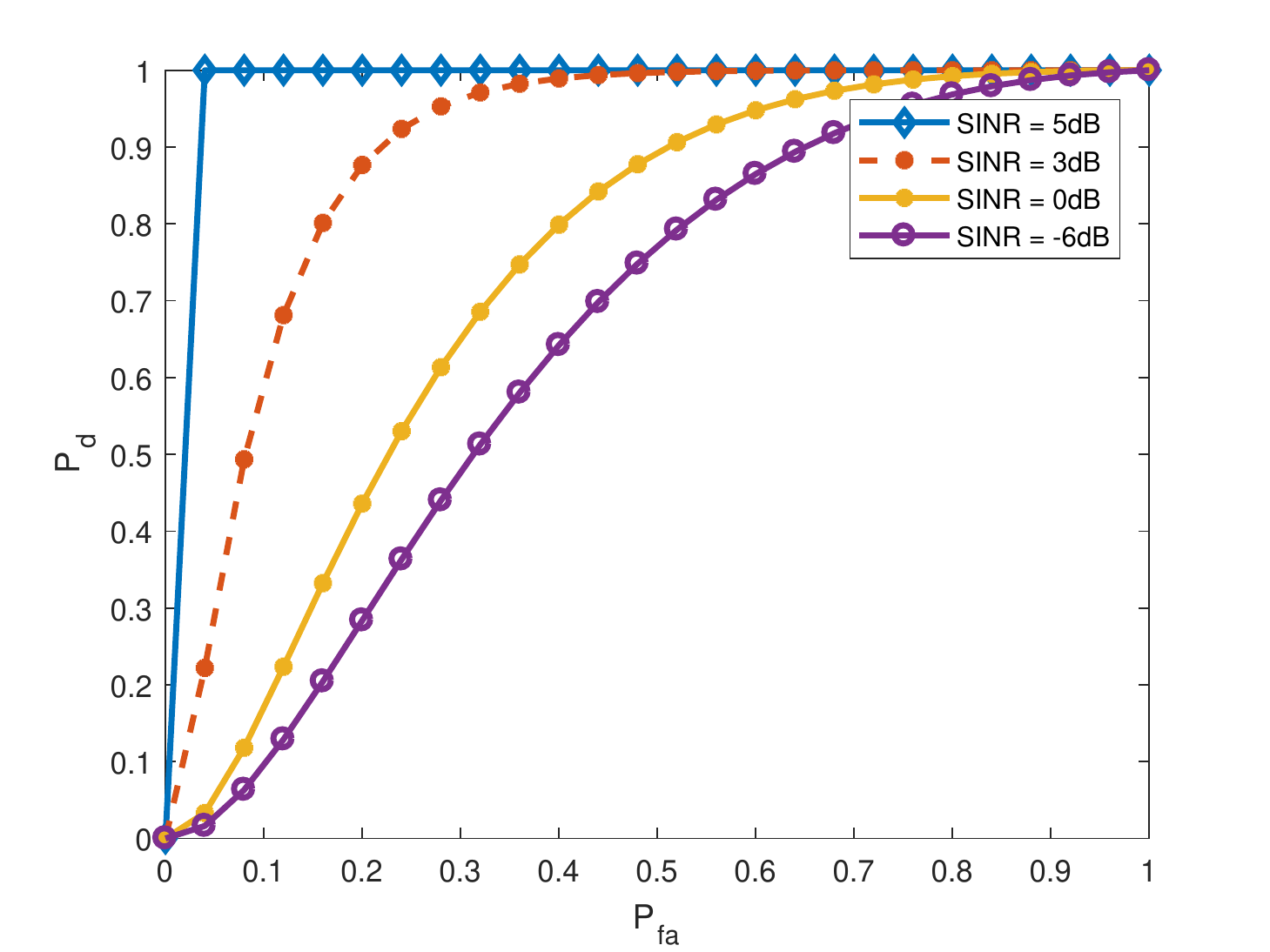}
\caption{ROC curves: for any pre-specified $P_{fa}$, $P_d$ increases with increase in SINR.}
\centering
\label{fig:ROC2}
\end{figure}

We define signal-to-interference-plus-noise ratio (SINR) at the reader as:  SINR = $\frac{\eta^2 P_R}{\sigma^2_{RT}}$.\\
Fig. \ref{fig:ROC2} shows the receiver operating characteristic (ROC) curves for various different values of SINR. Note that $P_d = 1- P_{md}$ is the probability of correct detection. To obtain this plot, we sweep the $P_{fa}$ from zero to one and then the threshold is calculated accordingly, then, for the given threshold we compute $P_d$. We observe that for a give value of $P_{fa}$, $P_d$ increases with increase in the SINR. 

\begin{figure}[h!]
\centering
\includegraphics[width=3.7in]{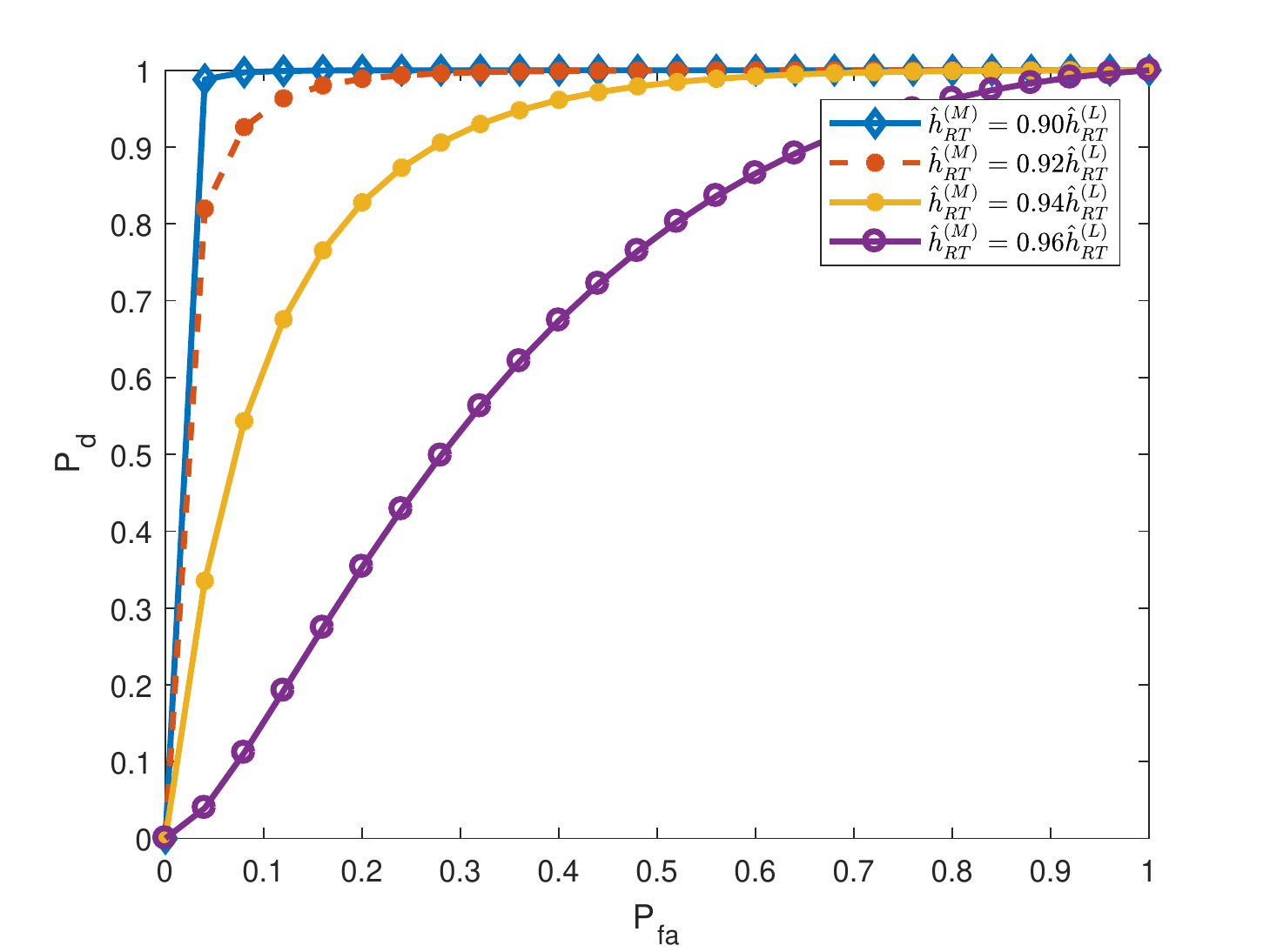}
\caption{ROC curves: for any pre-specified $P_{fa}$, $P_d$ increases as the fingerprint of the M-tag becomes more dissimilar to the fingerprint of the L-tag.}
\centering
\label{fig:ROC1}
\end{figure}

Fig. \ref{fig:ROC1} shows the ROC curves for different values of Eve's fingerprint. We observe that for any pre-specified $P_{fa}$, $P_d$ increases as the fingerprint $\hat{h}_{RE}$ of the M-tag becomes more dissimilar to the fingerprint $\hat{h}_{RT}$ of the L-tag. We fixed SINR to $5$ dB to obtain this result.

\section{Conclusion}

This work studied identity attack on an RFID-based backscatter communication system and proposed to utilize the so-called residual channel as the fingerprint of the tag(s) in order to verify the identity of the tag(s) at the reader. In our proposed scheme, the reader acquired the raw data via two-way (challenge-response) message exchange mechanism, did least-squares estimation to extract the fingerprint, and did binary hypothesis testing to do the authentication. We also provided closed-form expressions for the two error probabilities  of interest (i.e., false alarm and missed detection).  

Future work will look into other relevant features/fingerprints (e.g., tag antenna impedance, tag antenna gain etc.) which could be used as device fingerprints for authentication in near-field RFID systems.

\footnotesize{
\bibliographystyle{IEEEtran}
\bibliography{main}
}
\vfill\break
\end{document}